# Butterfly effect and spatial structure of information spreading in a chaotic cellular automaton

Shu-Wei Liu,[1,2] J. Willsher,[3] T. Bilitewski,[4] Jin-Jie Li,[2] A. Smith,[5] K. Christensen,[2,6] R. Moessner,[1] and J. Knolle[3,7,2]

[1]*Max-Planck-Institut für Physik komplexer Systeme, Nöthnitzer Straße 38, 01187 Dresden, Germany*
[2]*Blackett Laboratory, Imperial College London, London SW7 2AZ, United Kingdom*
[3]*Department of Physics TQM, Technische Universität München, James-Franck-Straße 1, 85748 Garching, Germany*
[4]*Center for Theory of Quantum Matter, University of Colorado, Boulder, Colorado, 80309, USA*
[5]*School of Physics and Astronomy, University of Nottingham, Nottingham NG7 2RD, United Kingdom*
[6]*Centre for Complexity Science, Imperial College London, London SW7 2AZ, United Kingdom*
[7]*Munich Center for Quantum Science and Technology, 80799 Munich, Germany*

Inspired by recent developments in the study of chaos in many-body systems, we construct a measure of local information spreading for a stochastic cellular automaton in the form of a spatiotemporally resolved Hamming distance. This decorrelator is a classical version of an out-of-time-order correlator studied in the context of quantum many-body systems. Focusing on the one-dimensional Kauffman cellular automaton, we extract the scaling form of our decorrelator with an associated butterfly velocity $v_b$ and a velocity-dependent Lyapunov exponent $\lambda(v)$. The existence of the latter is not a given in a discrete classical system. Second, we account for the behavior of the decorrelator in a framework based solely on the boundary of the information spreading, including an effective boundary random walk model yielding the full functional form of the decorrelator. In particular, we obtain analytic results for $v_b$ and the exponent $\beta$ in the scaling ansatz $\lambda(v) \sim \mu(v - v_b)^\beta$, which is usually only obtained numerically. Finally, a full scaling collapse establishes the decorrelator as a unifying diagnostic of information spreading.

## I. INTRODUCTION

A central hallmark of chaotic systems [1] is their sensitivity to perturbations: even a small change in initial conditions leads to entirely unpredictable, and large, differences in the state of the system at later times. This is popularly captured by the "butterfly effect" [2], which in many-body systems encodes two distinct notions: first, the exponential growth of the perturbation, giving rise to the notion of the Lyapunov exponent $\lambda$, characterizing the growth *with time* [3], and, second, information spreading *in space*, whereby the "effect" of the butterfly's wingbeat is felt at a distant location only with a time delay given by the "ballistic" propagation speed known as butterfly velocity $v_b$ [4,5].

A prominent recent line of investigation of chaos in quantum many-body systems revolves around the study of information scrambling, where out-of-time-order correlators (OTOCs) have been employed to measure the propagation of quantum chaos [6]. OTOCs can be understood as two-time correlation functions in which operators are not chronologically ordered and are a simple measure of the "footprint" of an operator that spreads in space [7], thus naturally measuring the spread of information. In chaotic systems, this quantity may grow exponentially in time, governed by the Lyapunov exponent $\lambda$ [8]. Recently, OTOCs have been studied extensively as an early- to intermediate-time diagnostic of quantum chaos or information spreading in a number of different quantum models whose time evolution can be generated by different dynamics such as Floquet dynamics [9,10], random unitary circuits [11–14], and time-independent Hamiltonians such as integrable spin chains [15,16], generalized Sachdev–Ye–Kitaev models [17], diffusive metals [18], and Luttinger liquids [19].

In that context, analogs of OTOCs were recently developed for classical systems [20–26]. For example, for spin chains the decorrelator $D(x, t) = 1 - \langle \mathbf{S}^A(x, t) \cdot \mathbf{S}^B(x, t) \rangle$ between two copies of spin configurations $\mathbf{S}^{A/B}$, which at $t = 0$ differs locally by only a small spin rotation, is a semiclassical version of an OTOC [20]. For Heisenberg magnets, this exhibits ballistic propagation with a light-cone structure governed by a butterfly velocity in the high-temperature regime without long-range magnetic order but with spin diffusion [20].

A common feature observed in classical and quantum models is the exponential growth (or decay) of the OTOCs (analogs) along rays of constant velocity $v = \frac{dx}{dt}$ quantified by velocity-dependent Lyapunov exponents (VDLEs) [8]. For intermediate/late times it takes the scaling form

$$D(x, t) \propto e^{-\mu(v-v_b)^\beta t} = e^{-\lambda(v)t}. \quad (1)$$

Intriguingly, a unifying framework of VDLEs with $\lambda(v) = \mu(v - v_b)^\beta$ captures the spatiotemporal structure of information spreading in many-body quantum, semiclassical, and classical chaotic systems [1,3,5,20,27].

In this work, we provide a basic description of such information spreading in a minimal chaotic setting: the Kauffman cellular automaton (KCA). Our system choice is motivated by the fact that KCAs are minimal chaotic many-body models [28] which display a rich phenomenology, including a phase transition as a function of a tuning parameter, a probability $p$. They exhibit universal scaling, e.g., of the directed percolation universality class [29], and lend themselves to analytical insight [30]. Additionally, KCA have a wide range of applicability [30–36]: initially introduced to study fitness landscapes of biological systems and gene expression [37], they now appear also in optimization problems [38], random mapping models [39], and, most pertinently, the emergence of chaos [40].

Surprisingly, despite decades of research on chaotic CAs (see Ref. [41] for a review), the dynamics of chaos has been investigated only in terms of the global Hamming distance, and a local diagnostic has thus far been missing. Here, we construct an OTOC analog for KCA and explore the VDLE phenomenology. This analog enables us to uncover the ballistic spatiotemporal structure of perturbation spreading in the chaotic phase [Fig. 1(b)], in contrast to the decay of such "damage" [42] spreading in the frozen phase [Fig. 1(a)]. We develop a full microscopic theory of the VDLE, recovering the functional form (1), including an analytical calculation of the exponent $\beta$. Thus, we provide the tools for describing the sensitivity of chaotic many-body systems to perturbations through the framework of VDLEs.

## II. KCA MODEL AND CLASSICAL OTOC ANALOG

We focus on a generic dissipative dynamical system called an $NK$ model. Concretely, a local KCA is a system of $N$ Boolean elements $\sigma(x, t) = \pm 1$ which evolve in discrete time steps through rules which depend upon each site and its $2K$ nearest neighbors in one dimension [41]. Our KCA system evolves under a set of (annealed) local rules $\{f_{x,t}\}$:

$$\sigma(x, t+1) = f_{x,t}[\sigma(x-K, t), \ldots, \sigma(x, t), \ldots, \sigma(x+K, t)], \quad (2)$$

which are random with probability $p$ in space and time:

$$f_{x,t} = \begin{cases} +1 & \text{with probability } p, \\ -1 & \text{with probability } 1-p. \end{cases} \quad (3)$$

Essentially, such local rules map $(2K + 1)$ inputs to a single output whose value is $+1$ or $-1$. At any particular $t$, the same local inputs always lead to the same output.

In a pioneering work, Derrida and Stauffer [43] showed that KCAs display a chaotic-to-frozen phase transition controlled by the parameter $p$ (see Fig. 1, inset). The two phases are distinguished by the decay or spread of localized perturbations diagnosed with the *global Hamming distance*,

$$H(t) = \frac{1}{2N} \left\langle \sum_x |\sigma^A(x, t) - \sigma^B(x, t)| \right\rangle_p, \quad (4)$$

between two copies of the system $\sigma^{A/B}(x, t)$ which differ by a single inverted site in the initial state at $t = 0$. This measure is then ensemble averaged over realizations with the same probability $p$. The distance grows linearly in the chaotic phase up to the physical boundary of the system and decays to zero in the frozen phase (see Fig. 1).

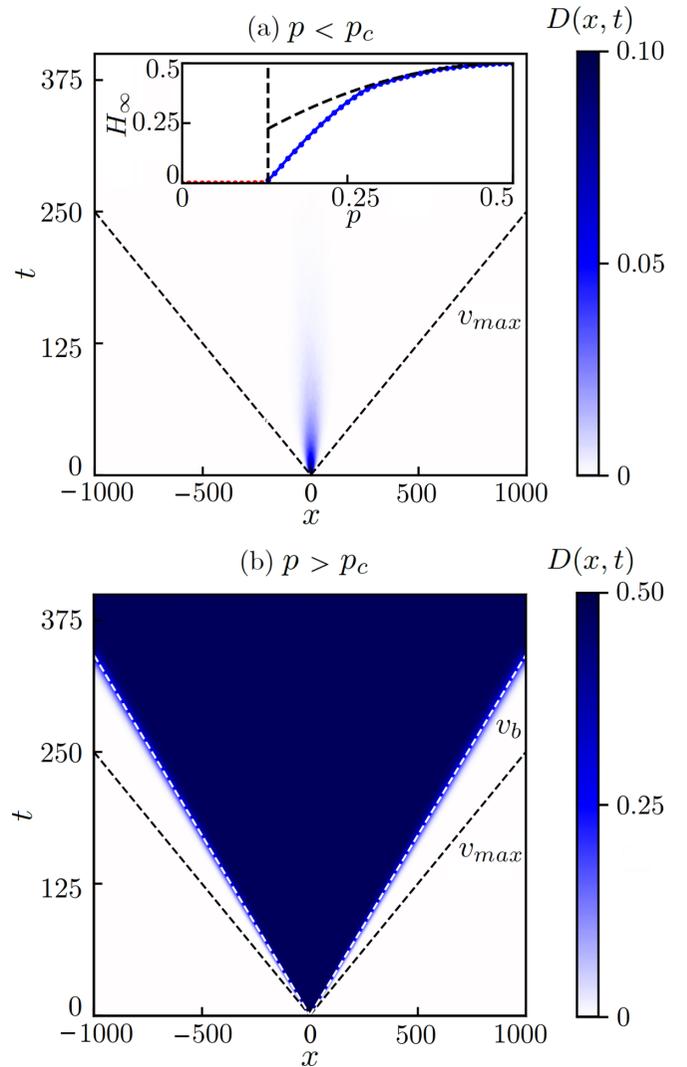

FIG. 1. Light-cone structure of the decorrelator $D(x, t)$ with $N = 2048$ and $K = 4$, with a single spin flip at the origin $x = 0$ when $t = 0$. (a) The frozen phase at $p = 0.11$ and (b) the chaotic phase at $p = 0.40$. In the chaotic phase, the two dashed lines are $v_b$ and $v_{\max}$, respectively. Inset: Mean-field prediction (dashed line) of the long-time Hamming distance $H_\infty = H(t \to \infty)$ as a function of $p$, compared with numerical data. $p_c = 0.13$. Close agreement is observed for $p > 0.27$.

Analogous to the classical Heisenberg chain [20], we take the classical OTOC as the *local* distance between two copies of the same system which differ by only a local perturbation of the initial conditions, thus leading to the *decorrelator*

$$D(x, t) = \tfrac{1}{2}[1 - \langle \sigma^A(x, t) \cdot \sigma^B(x, t) \rangle_p]. \quad (5)$$

It is nothing but a *local Hamming distance*, which is related to the global distance by $H(t) = \frac{1}{N} \sum_x D(x, t)$.

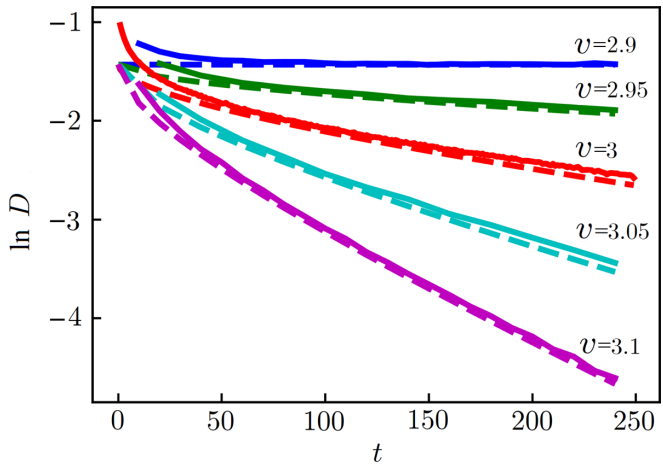

FIG. 2. The decorrelator $D(x, t)$ along rays of different velocities for $p = 0.4$, which is representative of all $p > p_c$. As $\ln D = -\lambda(v)t = -\mu(v - v_b)^\beta t$, the slopes represent $-\mu(v - v_b)^\beta$. Thus, $v_b = 2.9$ where the slope is zero. We compare the full $D(x, t)$ numerical data (solid lines) and the prediction (dashed lines) of the boundary random walk model.

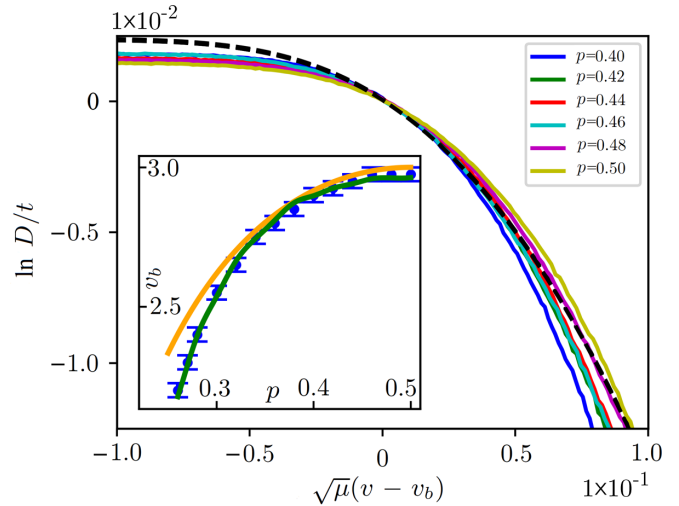

FIG. 3. Scaling collapse performed for a range of $p$ according to Eq. (14). At each specific $p$, the data are obtained by taking the value of $D$ in the $t \to \infty$ limit as permitted by computational power. The $x$ axis is rescaled to $\sqrt{\mu}(v - v_b)$, where both $\mu$ and $v_b$ are obtained by the slope method exemplified in Fig. 2, assuming $\beta = 2$ in the scaling form. The black dashed line is obtained by numerically evaluating $D(x, t)$ from Eq. (11) for $p = 0.5$. Inset: Comparison between $v_b$ obtained from the full $D(x, t)$ data (blue dots), from the boundary velocity (green line), and from the analytic calculation (8) (orange line).

## III. NUMERICAL RESULTS

In Fig. 1 we show the spatiotemporal evolution of the decorrelator $D(x, t)$ for representative values of $p$ below and above $p_c$. First, in the frozen phase $D(x, t)$ initially spreads but then decays in time and space to zero. This attenuation of the local decorrelator reflects the vanishing of the long-time value of the Hamming distance in the frozen phase. Second, in the chaotic phase $D(x, t)$ spreads with an apparent light-cone structure.

Because of the locality of KCA rules, the speed of the damage spreading measured with $D(x, t)$ is necessarily bounded by the maximum velocity $v_{\max} = K$, but the actual spread is slower, with the butterfly velocity $v_b < K$. To demonstrate the presence of $v_b$, we plot the behavior of the decorrelator along rays of constant velocity. In Fig. 2 we see that, after a transient effect, there is a ray (blue lines, corresponding to $v = 2.9$) along which $D(x = v_b t, t)$ is constant for $t > 100$. This establishes the presence of a butterfly velocity $v_b < v_{\max}$ and is an efficient way of extracting $v_b$ as a function of $p$ (see blue dots in the inset of Fig. 3).

Next, we investigate whether such behavior follows the general VDLE phenomenology of Eq. (1). We note that in contrast to previous work on Heisenberg spin chains, the presence of Lyapunov exponents in KCA is far from obvious. In fact, for small distances one does not expect an exponential pickup as a function of time because the local perturbation between the two copies $\sigma^{A/B}$ is necessarily big because of the discrete and bounded nature of the inverted site. However, we find that for distances $x$ far away from the perturbation and after disorder averaging one may still observe an exponential pickup in time of the OTOC analog. Because of the discrete nature of the variables the scaling behavior appears in only a small window around $v_b$ as the decorrelator quickly saturates to $D = D_0$ for velocities $v < v_b$ and quickly decays to $D = 0$ for $v > v_b$.

Again, a clear picture emerges by studying $D(x, t)$ along "rays" of constant velocity $v = x/t$. As shown in Fig. 2, we observe an exponential decay of the OTOC in time, with a VDLE $\lambda(v)$. For rays with $v > v_b$, the linear decay of $\ln D(x = vt, t) = -\lambda(v)t$ in the long-time limit demonstrates that the decorrelator indeed decays exponentially with a VDLE $\lambda(v)$.

Overall, the numerical results establish the presence of a butterfly velocity $v_B$ and validate the VDLE framework in KCA systems at long times. This is a surprising feature given the discrete, and large, nature of "perturbations" in the Boolean network, as such a framework is usually observed for continuous and infinitesimal perturbations [8].

To quantitatively analyze the small *active* region around the wave front set by $v_b$ where the damage actively spreads, we define for each sample the boundary of a spreading perturbation as the farthest point from the center which differs from the unperturbed system. As shown in Fig. 4, the probability density of boundary positions $P(x)$ approaches a Gaussian profile in the long-time limit with a width $\sigma_x \propto \sqrt{t}$. We find that its mean is in quantitative agreement with the probability-dependent butterfly velocity $v_b(p)$ for all $p$ (see the inset in Fig. 3). The Gaussian behavior of the boundary therefore motivates a random-walk-like description of the active region.

## IV. BOUNDARY RANDOM WALK MODEL

We can now develop a microscopic statistical model of the boundary dynamics starting from the microscopic KCA rules. The basic ingredient for our random walk model is the expectation value of the outwards move of a damage site in each time step: The farthest the boundary could move outwards in one time step is $K$, with probability $p(K) = p_d = 2p(1-p)$; the probability of moving $x < K$ steps is $p(x) = p_s^{K-x} p_d$, where $p_s = p^2 + (1-p)^2$. From this we obtain the mean and variance of the boundary steps, and the central limit theorem (CLT) allows us to obtain the full distribution (valid in the long-time and $p \gg p_c$ limits) as a Gaussian.

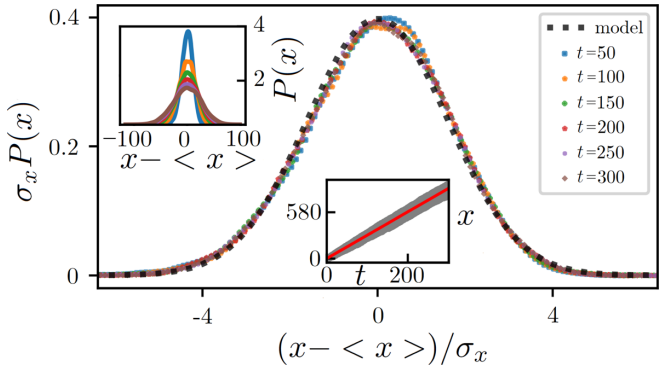

FIG. 4. Scaled probability density of boundary positions (defined in the text) of $10^4$ different initial configurations plotted against time. The black dotted line is the Gaussian profile predicted by the random walk model. The earliest two times ($t = 50, 100$) show slight deviations from the equilibrium/long-time limit. Top inset: Unscaled boundary probability density at $t = 50, 100, \ldots, 300$. Bottom inset: Raw data of boundary positions plotted against time. The red line is the average boundary, with the velocity $v = x/t = 2.9$.

This model correctly predicts various aspects of the decorrelator. First, we obtain the long-time value of $D(x,t)$ inside the light-cone as given by $D_0 = D(x, t \to \infty) = 2p(1-p)$, which corresponds to the spins pointing up and down with random probabilities $p$ and $1-p$. In the chaotic phase, $D_0$ is independent of $x$, and therefore, $H_\infty = D_0 = 2p(1-p)$ based on the sum rule [see the dashed line in the inset in Fig. 1(a)].

Second, the model predicts the butterfly velocity. The expectation of the outwards movement of the boundary in a single time step is approximated by[1]

$$\langle \Delta x \rangle = \sum_{x=-\infty}^{K} x p(x) = \sum_{x=-\infty}^{K} x p_s^{K-x} p_d, \quad (6)$$

so that the spatial profile as a function of time is $\langle x(t) \rangle = t \langle \Delta x \rangle$. By the CLT, one expects this profile to approach a Gaussian at large times, with a variance given by $\sigma^2(t) = t[\langle (\Delta x)^2 \rangle - \langle \Delta x \rangle^2]$. One may evaluate the moments of the probability density using the following manipulation:

$$\langle \Delta x^n \rangle = \sum_{x=-\infty}^{K} x^n p_s^{K-x} p_d = \left( \frac{d}{d \ln p_s} + K \right)^n \sum_{x=0}^{\infty} e^{\ln p_s x}. \quad (7)$$

This expression may be evaluated using the geometric series and gives a surprisingly simple expression for the butterfly velocity:

$$v_b(p) = K - \frac{p_s}{p_d} = (K-1) - \frac{(1-2p)^2}{2p(1-p)}. \quad (8)$$

It agrees with the full model's butterfly velocity at large $p$ away from $p_c$, as shown in the inset in Fig. 3.

Third, the standard deviation of the boundary distribution is

$$\sigma^2(t) = t\left( \frac{p_s}{p_d} + \frac{p_s^2}{p_d^2} \right) \equiv \frac{t}{2\mu^2(p)}, \quad (9)$$

which confirms the Gaussian form of the boundary random walk with a variance that scales linearly with time, as shown in Fig. 4.

Fourth, the cumulative distribution of the boundary then allows us to obtain the complete functional form of the OTOC as a function of $v$, governed by its inverse width $\mu(p)$. For any one realization, the boundary will trace a biased random walk in time, and inside of its chaos spreading will be its own scrambling region with the expected value $D_0$ for the OTOC. The OTOC therefore acts like the cumulative probability density function of the boundary's probability density, weighted such that it has a central value of $D_0 = 2p(1-p)$:

$$D(x,t) = D_0 \int_x^\infty G(x',t) \, dx', \quad (10)$$

where $G(x',t)$ is the Gaussian distribution. In our approximate model at high $p$ and late times we may take the Gaussian limit, therefore deriving the following form for the OTOC in terms of the error function $\text{erf}(x)$:

$$D(x,t) = \frac{D_0(p)}{2}\left[ 1 - \text{erf}\left( \frac{x - v_b(p)t}{\sqrt{2}\sigma(t)} \right) \right]$$
$$= p(1-p)[1 - \text{erf}([v - v_b(p)]\sqrt{\mu(p)t})], \quad (11)$$

where we have used $x = vt$ and $\sigma^2(t) = t/2\mu^2(p)$. By taking the series expansion of the error function in $v - v_b$ for large $x$ we recover an exponential decay of $D(x,t)$; in logarithmic form this is (when $v > v_b$)

$$\ln D(v,t) = \ln p(1-p) - \tfrac{1}{2} \ln \mu(v - v_b)^2 \pi t - \mu(v - v_b)^2 t. \quad (12)$$

Comparing this result to the general scaling forms predicted by previous works on spin chains,

$$\ln D(v,t) \sim -\mu(v - v_b)^\beta t, \quad (13)$$

one may identify the same behavior in the long-time limit. Figure 2 shows the predicted analytic form of the decorrelator (dashed lines) compared to the numerical data (solid lines). This quantitative agreement in the long-time limit confirms that the boundary controls the dynamics of chaos spreading of the full KCA model.

Most notably, in the long-time limit the analytical model recovers the linear decay of the decorrelator in time, described by a VDLE for $v$ close to $v_b$:

$$\ln D(x = vt, t) = -\lambda(v)t = -\mu(p)(v - v_b)^2 t, \quad (14)$$

with an exponent $\beta = 2$. Therefore, in this regime we expect a data collapse around $v_b$ by plotting $\ln(D)/t$ against $\sqrt{\mu(p)}(v - v_b)$. Indeed, as shown in Fig. 3, the data of these two variables for a range of probabilities fall onto the single curve of the analytical prediction (black dashed line).

---

[1]The use of $-\infty$ as the lower limit of the sum is a useful approximation for large $K$ as it provides a simple closed-form expression for $v_b$ and $\sigma$.

## V. DISCUSSION

Given the discrete nature of the dynamics, the quantitative agreement between our model and the full numerical simulation is somewhat surprising but highlights the universal features of chaos spreading. Crucially, our analysis is valid only after sample averaging and in the long-time limit, where the effects of the discrete perturbation and dynamics have been smoothed out. In particular, when approaching the critical point $p_c$ from above, we see systematic deviations due to fluctuations.

Our work on a minimal classical model was inspired by recent developments in the study of chaos in quantum many-body systems where OTOCs have become a powerful quantitative tool. One important prediction in that context is that the butterfly velocity is bounded at high temperatures $v_b(T \to \infty) \sim v_{\text{LR}}$ [44], where $v_{\text{LR}}$ is the Lieb-Robinson velocity, which is the upper limit of information propagation in short-range interacting nonrelativistic quantum systems [27,45]. For KCA one may connect the model parameter $p$ to a temperature $T$ via $p = e^{-1/T}/(e^{1/T} + e^{-1/T})$ [46]. Then from Eq. (8) we can obtain the full temperature dependence of the butterfly velocity,

$$v_b(T) = (K - 1) - \sum_{k \text{ even}} \frac{2^k}{k!} \frac{1}{T^k} \quad (15)$$

$$= (K - 1) - \frac{2}{T^2} - \frac{2}{3T^4} - \cdots, \quad (16)$$

and the high-temperature limit is $v_b(T \to \infty) = K - 1$. This expression confirms that also in our classical many-body system the maximum velocity of information spreading is always less than the maximum $v_{\max} = K$ allowed by the local dynamics.

## VI. CONCLUSION AND OUTLOOK

We have constructed a local diagnostic of information spreading for a one-dimensional random CA in analogy with recent semiclassical versions of OTOCs. We demonstrated that it displays ballistic propagation characterized by a butterfly velocity and exponential growth in time captured by a VDLE. We developed a random walk model of the boundary of information spreading which permits the calculation of the full functional form of the classical OTOCs, including the exponent $\beta$ of the VDLE.

An obvious extension of our work is to consider the two-dimensional KCA where an even richer phenomenology is expected. Even though our boundary random walk model proposed here is simple, it has many physical and mathematical aspects that remain unexplored. For example, if the random walk is approximated to have a continuous walk distance which still follows the same probability distribution, can it have a closed-form solution for early times without having to invoke the CLT? Alternatively, it would be worthwhile to explore a Langevin-like description of the boundary, which is particularly useful when considering KCA in higher dimensions. For example, in two-dimensional KCAs, we expect a diffusionlike process for the perturbation which could potentially percolate across the lattice. In particular, it is interesting to investigate whether this falls into any percolation universality class where established critical exponents could be linked to the exponent of the classical OTOCs.

Generally, we expect our decorrelator and theoretical tools to be applicable to other stochastic models with discrete variables, which are widely used to describe the dynamics of both quantum and classical many-body systems. While, here, we use a classical random walk model to understand the decorrelator in KCAs, we note that recently, quantum random walk techniques have resulted in improved bounds on the growth of OTOCs in large-$S$ models [47]. We expect that different variants of the boundary theory developed here should be able to predict the scaling forms in these systems as well. In particular, it would be interesting to see them applied to models with charge or dipole conservation rules [48,49] in which conjectured critical exponents could potentially be derived analytically.

## ACKNOWLEDGMENTS

We thank S. Bhattacharjee for useful discussions, specifically for collaboration on decorrelators in Heisenberg models. This work was, in part, supported by the Deutsche Forschungsgemeinschaft under SFB 1143 (Project-id 7247310070) and cluster of excellence ct.qmat (EXC 2147, Project No. 390858490). A.S. was supported by a Research Fellowship from the Royal Commission for the Exhibition of 1851.